# An Approach to Find Missing Values in Medical Datasets

B.Mathura Bai
Faculty of Information Technology
VNR VJIET
Hyderabad
INDIA
mathura.aniketh@gmail.com

N.Mangathayaru
Professor
Dept. of Information Technology
VNR VJIET, Hyderabad
INDIA
manga.surya@gmail.com

B.Padmaja Rani
Professor
Computer Science & Engg Dept.
JNT University, Hyderabad
INDIA
padmaja_jntuh@yahoo.co.in

## ABSTRACT

Mining medical datasets is a challenging problem before data mining researchers as these datasets have several hidden challenges compared to conventional datasets. Starting from the collection of samples through field experiments and clinical trials to performing classification, there are numerous challenges at every stage in the mining process. The preprocessing phase in the mining process itself is a challenging issue when, we work on medical datasets. One of the prime challenges in mining medical datasets is handling missing values which is part of preprocessing phase. In this paper, we address the issue of handling missing values in medical dataset consisting of categorical attribute values. The main contribution of this research is to use the proposed imputation measure to estimate and fix the missing values. We discuss a case study to demonstrate the working of proposed measure.

## Categories and Subject Descriptors:

I.2.6 Learning –Knowledge acquisition. H.2.8 Database applications: Data mining I.2.6 Artificial Intelligence: Learning - parameter learning

## 1. INTRODUCTION

Disease Prediction and Classification is evolving as an important research problem in the fields of medical and health informatics.

Medical data has several hidden challenges when compared to conventional datasets. The first step is the sample data collection which needs defining the attributes which functionally define the target specification. This phase of sample data collection is not free from challenges. The possibility of the presence of missing values for some attributes itself initiates the first challenge before the data mining researchers.

One simple method to handle the missing values is to simply eliminate the medical record consisting of missing values. But this does not solve the problem. This is because, this medical record might contain the values of attributes which might be the deciding factors in predicting the diseases in some situations. If we do not eliminate such records consisting missing values, then the next strategy is to consider those records for the purpose of training. In such cases, we must fix the missing values. This is carried by several researchers in different ways.

The most common way is using the distance measure to perform imputation of these missing values and replace these missing values using the suitable values. This requires proper estimation to be made. This is because any failure to do so would give fault outcome.

The problem of mining medical records may be classified into the problem of handling the missing values, performing classification and clustering process which requires that the class label of each record is available. This is a supervised learning process. However clustering medical data to group similar medical records or patients is an un-supervised process of learning.

Most works in the literature make use of Euclidean distance measure to fix the missing values. Some researchers try to cluster the medical data and then fix the missing values using k-means algorithm for clustering. Our findings indicates that there is a huge scope for research towards fixing missing values by designing new distance measures which can satisfy the basic properties of distance measures. If this is done then, we can estimate the missing values of the records containing the class label. In this paper, the section-2 discusses the related works. Section-3 points out the research issues in medical data mining. Section-4 introduces the problem definition and a generalized approach for handling the medical datasets.

## 2. RELATED WORKS

Missing Values are most common when various field experiments and clinical trials are carried out. These missing values throw hidden challenges to the data miner and analyst. In this section we, discuss various research works carried out recently in healthcare and medical data mining. In [1], the authors discuss top-10 data mining algorithms which are most promising. In [2, 3, 4, and 5] the authors work involves predicting the heart disease using data mining approaches.

The work in [6] deals with the knowledge discovery issues in data mining. The work in [7] deals with evolutionary approach in data mining and [8] deals with finding interesting rules from the dataset. In [9], the authors discuss the application of frequent items finding algorithm to find the periodical frequent patterns from transaction data.

The authors [10, 14, and 15] propose an approach for handling the missing values when the feature values are continuous in nature using non-parametric discretization technique. They try to impute the missing values. They use the concept of z-score to handle the missing values in the medical records. This requires the underlying data to be continuous.

In [11, 13] the authors use the concept of imputation to handle the missing values considering dengue fever dataset. They design the procedure to impute the missing attribute values for continuous attributes which may be numeric and categorical. They also use the decision tree for generating efficient decision rules for classification.

In [12], the authors compare the various approaches available in the literature for handling the missing values with the benchmark datasets. In [17, 18] the authors perform classification of medical records. The research in [26-40] includes contributions of various researchers towards finding missing values, classification, prediction, clustering medical records.

## 3. Research Challenges

### 3.1 Dimensionality Reduction

The process of dimensionality reduction w.r.t medical datasets requires handling the data without any loss. This is mainly because unlike the conventional approach where we try to reduce the dataset by eliminating irrelevant , least important features of the dataset through preprocessing which includes stop word elimination, removing stemming words, this approach does not hold good for the case of medical datasets.

Dimensionality reduction is the major concern when dealing with the large datasets and very large databases. However in the view of mining medical records, this problem becomes much more complex. This is because; we must perform dimensionality reduction but at the same time we must make sure that we retain all the information without any loss which is not as simple as we view it to be.

Dimensionality reduction techniques have been discussed extensively in various research works in the literature. The preprocessing phase also requires normalizing the data to make it feasible for handling efficiently. Once the data is normalized, the next problem is with the missing values.

### 3.2 Missing Values

A peculiar problem which arises when handling medical datasets is the presence of missing values in the dataset which is very natural. This is because the data may have been collected; sampled from various clinical trials, field experiments or any other traditional mechanism. Proper care must be taken to handle missing values suitably and accurately.

A simple approach to handle missing values would be to discard the whole record which essentially contains the missing value of an attribute. But, this do not solve the problem as, we are missing this record. There is a chance that this record may be important and deciding factor when performing prediction and classification.

The problem of handling missing values has been widely dealt by data mining practitioners and researchers by exploring various approaches to handle the missing values [26-33]. Handling missing values requires the designing suitable imputation measures or using the existing measures. In case, we design new measure, it should satisfy the typical properties of a distance measure. There is a scope for research in this direction. The researchers may come up with suitable measures or alternately validate the use of mathematical functions to suit the need for fixing and filling missing values.

### 3.3 Functional Relation among Attributes

Another important challenge and problem which cannot be left unexplored is discovering the functional relation among the attributes of the medical datasets. This must handled very carefully as the features or attributes are the major elements.

### 3.4 Imputation Measure

The choice of suitable imputation mechanism, approach and measure is critical for handling medical data. Estimation of missing values using an effective imputation measure shall be the deciding stage when performing classification and prediction. The training set, number of samples is considered critical when handling medical datasets.

One of the approaches to handle missing values is the use of proper imputation measure which can efficiently fix the missing values. Alternately, we may design a suitable imputation measure which may be used to estimate the missing value, and replace that specific missing value by an equivalent value computed using this imputation measure. Some of the works in this direction includes [10, 14].

### 3.5 Deciding on the Number of Attributes
One of the challenging problems when performing data mining is the number of attributes. As the number of attributes, increases, the complexity of handling the dataset also increases. This is because of the increasing attribute combination which makes the problem analysis NP-Hard. The complexity shows non-linearity.

Dimensionality reduction techniques such as feature selection, reduction, PCA, SVD are usually applied to handle when the attributes are very large. In some situations, domain experts and domain knowledge may help reduce the attributes. The domain knowledge may also be useful to obtain the dependency relation among the attributes. Parameter tuning is also one of the problems when handling the medical databases and datasets.

### 3.6 Update Anomaly
Medical data is not static. It is time variant. It is multi-dimensional. Since the medical data keeps updating from the readings arriving from laboratory tests, the data mining approaches, methods, methodologies, algorithms designed should be able to dynamically learn, build and update the knowledge database incrementally.

Incremental Data Mining approaches are better suitable for handling medical databases as against to static methods and approaches. Techniques such as, incremental frequent pattern mining algorithms, incremental clustering, Dynamic Item set finding algorithms, handle the time varying nature of medical datasets.

### 3.7 Problem with Data Interpretation
The medical data usually comes from various homogeneous and heterogeneous sources which bring the problems such as the data inconsistency, data incompleteness, data interpretation. This is because; various sources may consider different terminologies, standards, attributes. The attributes may be continuous or not.

### 3.8 Imbalanced Nature of Medical Datasets
Medical datasets are usually imbalanced. In such a case, the dataset must be preprocessed, before using the same. The use of preprocessing techniques available must be done suitably to handle such anomalies.

## 4. Problem Definition
Given a dataset of medical records, the problem is to classify the medical records to one of the classes labeled. Applying data mining techniques to medical datasets helps in identifying the hidden knowledge which helps in prediction and classification. There are various approaches for classification and clustering the datasets. Each algorithm has its own pros and cons. Also, the dataset over which we apply these techniques may not be free from missing values as all attribute values for every record corresponding to a schema may not be recorded and as a result of this, every value may not be present. The first step in handling the dataset may itself be challenging because this step requires handling missing values.

There is scope for research to design algorithms for handling missing values in the medical records. Once we handle the missing values in the medical dataset, the dataset M may be transformed to M'. This M' must then be normalized to make the medical dataset feasible for performing classification using the existing algorithms. The third challenge is in identifying the outlier attributes or features and then reducing this M' to M''. This process is called as dimensionality reduction.

Finally, we may classify the existing record in to one of the class labels available. To handle missing values, we must know the corresponding class labels of each record. Otherwise this process becomes much more complex. All the existing algorithms handling missing values put a constraint that the class labels need to be known Apriori using which the missing values shall be filled. In case, the class labels are missing, we must first try to identify the nearest class label and then fill the missing values in the medical records.

## 5. PROPOSED MEASURE

### 5.1 IMPUTATION ALGORITHM
**Step1:**

Place the dataset in record and attribute matrix format with records as rows and attributes as columns

**Step2:**

Decompose the medical dataset into two parts. The first part consists of records with no missing values and second part consisting records with missing values.

**Step3:**

Normalize the datasets obtained in step-2 suitably. This involves mapping the attribute values to suit the need for applying the proposed measure to find missing values i.e we fix the domain of attribute values.

**Step-4:**

Choose one record with missing attribute value or missing values, each time to fix the missing values of the record.

**Step-5:**

From the record considered in above step, discard the attribute column consisting missing values (second part of dataset). Similarly, discard this attribute column when considered remaining records with no missing values (first part of dataset)

**Step-6:**

Apply the proposed measure to find the similarity between the new record and existing records without missing values.

**Step-7:**

The class label of new test record is the label of the record to which the new test record shows maximum similarity.

**Step-8:**

Fill the missing attribute value of the test record with the corresponding attribute value of the record to which it has maximum similarity. We may also consider the frequency of occurrence in case of ambiguity.

**Step-9:**

Repeat the above steps for each record having missing values accordingly.

## 5.2 IMPUTATION MEASURE

In this section, we discuss the proposed imputation measure which is used to fix the missing attribute values present in the medical records.

The Proposed Imputation measure is defined as given by Equation. 1

$$IMMV = \frac{1}{2} * \left[ 1 + \frac{\sum_{k=1}^{k=m} Sim(R_{ik}, R_{jk})}{total\ number\ of\ attributes} \right] \quad (1)$$

where

$$Sim(R_{ik}, R_{jk}) = 0.5 * \left[ 1 + e^{-\left(\frac{v_{ik} - v_{jk}}{std.dev(k)}\right)^2} \right] \quad (2)$$

The parameters $V_{ik}$, $V_{jk}$ denotes attribute values of $k^{th}$ attribute w.r.t records, $R_i$, $R_j$.

Similarly, $std.dev(k)$ denotes the Standard deviation of $k^{th}$ attribute w.r.t all records of dataset discarding missing value attributes.

The distance function is represented as given by

Distance, $\delta$ = 1-IMMV     (3)

## 6. CASE STUDY

Consider the Table.1 which has nine medical records and two records with missing values as shown below

Table 1. Sample Dataset, $M_D$

|       | $A_1$    | $A_2$ | $A_3$    | $A_4$ |
|-------|----------|-------|----------|-------|
| $R_1$ | $a_{11}$ | 5     | $a_{31}$ | 10    |
| $R_2$ | $a_{13}$ | 7     | $a_{31}$ | 5     |
| $R_3$ | $a_{11}$ | 7     | ?        | 7     |
| $R_4$ | $a_{12}$ | 5     | $a_{31}$ | 10    |
| $R_5$ | $a_{13}$ | 3     | $a_{32}$ | ?     |
| $R_6$ | $a_{12}$ | 9     | $a_{31}$ | 10    |
| $R_7$ | $a_{11}$ | 5     | $a_{32}$ | 3     |
| $R_8$ | $a_{13}$ | 6     | $a_{32}$ | 7     |
| $R_9$ | $a_{12}$ | 6     | $a_{32}$ | 10    |

Table.2 shows the records $R_1$, $R_2$, $R_4$, $R_6$, $R_7$, $R_8$, $R_9$ which have all the attribute values defined. We can observe that there are no missing values.

Table 2. Records with no missing values

|       | $A_1$    | $A_2$ | $A_3$    | $A_4$ |
|-------|----------|-------|----------|-------|
| $R_1$ | $a_{11}$ | 5     | $a_{31}$ | 10    |
| $R_2$ | $a_{13}$ | 7     | $a_{31}$ | 5     |
| $R_4$ | $a_{12}$ | 5     | $a_{31}$ | 10    |
| $R_6$ | $a_{12}$ | 9     | $a_{31}$ | 10    |
| $R_7$ | $a_{11}$ | 5     | $a_{32}$ | 3     |
| $R_8$ | $a_{13}$ | 6     | $a_{32}$ | 7     |
| $R_9$ | $a_{12}$ | 6     | $a_{32}$ | 10    |

Table.3 shows the records $R_3$, $R_5$. We can observe that there are missing values in these records w.r.t attributes $A_3$ and $A_4$ in records $R_3$, $R_5$ respectively.

Table 3. Records with Missing Values

|       | $A_1$    | $A_2$ | $A_3$    | $A_4$ |
|-------|----------|-------|----------|-------|
| $R_3$ | $a_{11}$ | 7     | ?        | 7     |
| $R_5$ | $a_{13}$ | 3     | $a_{32}$ | ?     |

Table 4. Standard.deviation of attributes

| $A_1$ | $A_2$ | $A_3$ | $A_4$ |
|---|---|---|---|
| 0.816497 | 1.46385 | 0.534522 | 2.91139 |

In Table.4, the standard deviation of all attributes w.r.t the records of Table.1 are computed and recorded. The attributes values for $A_1$ and $A_3$ are mapped to a different domain to make it suitable for processing by the algorithm and proposed measure. Similarly, Table.8 shows mapping of attribute values for records $R_5$ and $R_3$.

Table 5. Mapping $A_1$ and $A_3$ attribute values

|  | $A_1$ | $A_2$ | $A_3$ | $A_4$ |
|---|---|---|---|---|
| $R_1$ | 1 | 5 | 1 | 10 |
| $R_2$ | 3 | 7 | 1 | 5 |
| $R_4$ | 2 | 5 | 1 | 10 |
| $R_6$ | 2 | 9 | 1 | 10 |
| $R_7$ | 1 | 5 | 2 | 3 |
| $R_8$ | 3 | 6 | 2 | 7 |
| $R_9$ | 2 | 6 | 2 | 10 |
| Std.dev | 0.8164 | 1.4638 | 0.5345 | 2.9113 |

Table 6. Similarity of Record, R3 with Record in Table.2

|  | Similarity with $R_3$ |
|---|---|
| $R_1$ | 0.750079 |
| $R_2$ | 0.771048 |
| $R_4$ | 0.6206 |
| $R_6$ | 0.6206 |
| $R_7$ | 0.717678 |
| $R_8$ | **0.771595** |
| $R_9$ | 0.699342 |

Table 7. Similarity of Record, $R_5$ with Record in Table.3

|  | Similarity with , $R_5$ |
|---|---|
| $R_1$ | 0.531219 |
| $R_2$ | 0.671795 |
| $R_4$ | 0.567994 |
| $R_6$ | 0.542221 |
| $R_7$ | 0.692853 |
| $R_8$ | **0.835833** |
| $R_9$ | 0.706354 |

Table 8. Mapping Records with Missing Values

|  | $A_1$ | $A_2$ | $A_3$ | $A_4$ |
|---|---|---|---|---|
| $R_3$ | 1 | 7 | ? | 7 |
| $R_5$ | 3 | 3 | 2 | ? |

The similarity values of records $R_3$, $R_5$ w.r.t records $R_1$, $R_2$, $R_4$, $R_6$, $R_7$, $R_8$, $R_9$ is shown in Table.6 and Table.7 respectively. Table.10 shows the attribute values of records after fixing missing values using the proposed measure.

Table 9. Records with Mapping Values fixed

|  | $A_1$ | $A_2$ | $A_3$ | $A_4$ |
|---|---|---|---|---|
| $R_3$ | 1 | 7 | 2 | 7 |
| $R_5$ | 3 | 3 | 2 | 7 |

Table 10. Records with Missing Values fixed

|  | $A_1$ | $A_2$ | $A_3$ | $A_4$ |
|---|---|---|---|---|
| $R_3$ | 1 | 7 | **$a_{32}$** | 7 |
| $R_5$ | 3 | 3 | 2 | **7** |

The records $R_3$, $R_5$ show maximum similarity with the records of Table.6 and Table.7 respectively. Hence we fix the missing attribute values of these records w.r.t record $R_8$

## 7. CONCLUSIONS

Missing values are very common in medical data. Handling missing values is challenging and also an interesting area of research in perspective of data mining, information retrieval. In this work, we propose imputation measure which is used to fix missing values of attributes for the records of dataset. This is done by finding similarity values between records with missing values and records without missing values. A case study is discussed which shows procedure of fixing missing values.

## 8. ACKNOWLEDGEMENTS

We are thankful to anonymous reviewers who gave their valuable suggestions in bringing this work successfully. This work is not funded by any public, private or government agencies and has been carried as part of research work. We are also thankful to the Head of the Department, IT and Principal, VNR VJIET for motivating us towards research.